# Gravitational Effects of Weak Interactions at TeV Energies

C Sivaram and Kenath Arun

Indian Institute of Astrophysics, Bangalore

**Abstract:** Recently there has been a lot of interest in the search for extra dimensions. If gravity propagates in extra dimensions then gravity would become as strong as other interactions. This could also lead to the production of mini black holes. Here we have discussed how even without considering extra dimensions gravitational effects of weak interactions could show up at TeV energies.

There is a lot of interest in the search for extra dimensions at TeV energies, including at the LHC accelerator.[1] If gravity propagates in extra dimensions then $(4+n)$ gravity would become as strong as other interactions (such as electroweak) and unification would occur at $l_u \approx 10^{-17} cm$, corresponding to TeV energies.[2] This could also lead to the production of mini black holes with a radius $\approx l_u$, when we have a strong (unified) gravitational field (or coupling). Such black holes should decay on time scales $l_u/c \approx 10^{-27} s$ by Hawking radiation.[3]

However, even without considering extra dimensions it appears that gravitational effects of weak interactions, could show up at TeV energies.

For a region of radius r, the associated (short range) weak interaction energy is ($G_F$ is the Fermi constant):

$$E_W \approx G_F r^{-3} \qquad \ldots (1)$$

So the self gravitational energy is:

$$E_{GW} \approx \frac{G_F^2 r^{-6} G}{c^4 r} \approx \frac{G_F^2 r^{-7} G}{c^4} \qquad \ldots (2)$$

Where $G$ is the Newtonian gravitational constant. Note the $r^{-7}$ dependence!

For this to be a region of strong interaction, we have:[4]

$$E \approx E_{GW} \approx \frac{\hbar c}{r} \qquad \ldots (3)$$

($g^2/\hbar c \sim 1$ is the strong coupling, $e^2/\hbar c \sim \alpha$ is the electromagnetic coupling, and so on)

This would imply that the gravitational effects of the weak interactions would become strong when (from equations (2) and (3)):

$$r \approx \left(\frac{GG_F^2}{\hbar c^5}\right)^{1/6} \approx 10^{-19} cm \qquad \ldots (4)$$

This corresponds to an energy scale of around 100 TeV!

The above scaling suggests (that is $r^{-1/6}$ or $E^{1/6}$) that at energies ~10 TeV (like in the LHC), with the $E^{1/6}$ dependence of the gravitational-weak effects, the corresponding energies would be ~several MeV. This is testable. (That is the gravitational effects of the highly localised weak interactions at several TeV would amount to energies ~several MeV, causing measurable differences in particle energies, etc.)

Again two wave packets of extent given by equation (4) and separated by similar distance would gravitationally interact with energy:

$\sim \dfrac{G(G_F r^{-3})^2}{r} \propto r^{-7}$ again with similar strength. The interaction energy would also be ~several TeV.

However for a mini black hole to form (with the usual four dimensions) the energy required to be squeezed into the region of scale given by equation (4) is several orders more.[5]

The energy required for a region of extent $r$, to form a black hole is:

$$E_{bh} \approx \frac{c^4 r}{G} \qquad \ldots (6)$$

With equations (2) and (6), we get the required $r$ as:

$$r \approx \left(\frac{G^2 G_F^2}{c^8}\right)^{1/8} \qquad \ldots (7)$$

This gives $r \sim 10^{-24} cm$ for an energy of the particles of $\sim 10^6 - 10^7 TeV$! This could have consequence in astrophysics, especially for high energy cosmic rays which have energies of $\sim 10^{20} eV$.

Independently of extra dimensions, the extended uncertainty principle, if it has a weak fundamental length of $L_w$, could have testable effects at such TeV energies. In particular, equations (2)-(5) would get modified at lengths $< L_w$.

The generalised uncertainty principle (GUP) has the form:[6]

$$\Delta x \Delta p \geq \hbar + \frac{L_w^2}{\hbar}(\Delta p)^2 \qquad \ldots (8)$$

So for $\frac{\hbar}{\Delta p} >> L_w$, we just have the second term!

This implies that $\Delta x$ now increases with $\Delta p$. So equation (2) would be modified as:

$$E_{GW} \approx \frac{G_F^2 G}{c^4} r^{-7} \approx \frac{G_F^2 G}{c^4 \left(\frac{L_w}{\hbar/\Delta p}\right)^7} \approx \frac{G_F^2 G \hbar^7}{c^4 L_w^7}\left(\frac{c}{E}\right)^7 \approx \frac{G_F^2 G \hbar^7 c^3}{L_w^7 E^7} \qquad \ldots (9)$$

And $\Delta x$ increases with $\Delta p$ as:

$$\Delta x = \frac{L_w}{\left(\hbar/\Delta p\right)} \qquad \ldots (10)$$

If $L_w$ corresponds to $\sim 10^{-18} cm (10 TeV)$, equation (9) would suggest a sharp drop of the self gravitational energy with interaction energy at energies $>> 10 TeV$, in contrast to the earlier increase with energy.

This could reveal the existence of a fundamental weak length scale and a corresponding modified uncertainty principle at these energies as a consequence. These are all testable effects.

The modified phase space as implied by equations (8) and (9) would again have effects on particle decays. That is, we would have: $d^3 x d^3 p \geq \hbar^3 (1 + L_w^2 p^2)^3$, which could have drastic effects on decay times of unstable particles at these energies.[7]

The decay rate of the particles is given by:

$$d\Gamma = \frac{2\pi}{\hbar}|H_{fi}|^2 \frac{dN}{dW} \qquad \ldots (11)$$

Where the density of the final state is given by:

$$\frac{dN}{dW} = \frac{4\pi p^2 dp V}{(2\pi\hbar)^3} \qquad \ldots (12)$$

$W$ is the total energy of the final state given by: $W = \dfrac{p^2}{2m} + p_\nu c$

The density of the final state is:

$$\frac{dN}{dW} = \frac{(4\pi)^2 \left(W - \dfrac{p^2}{2m_e}\right) V^2 p_e^2 dp_e}{(2\pi\hbar)^6 c^3} \qquad \ldots (13)$$

The total rate is then given by:

$$d\Gamma = \frac{8g^2 |M|^2 p^7}{c^3 \hbar^7 m^2} \qquad \ldots (14)$$

With the modified phase space, the density of final state will be modified from that given by equation (12) to:

$$\frac{dN}{dW} = \frac{4\pi p^2 dp V}{(2\pi\hbar)^3 (1 + L_w^2 p^2)^3} \qquad \ldots (15)$$

Expanding the term in the denominator gives: $1 + 3L_w^2 p^2 + 9L_w^4 p^4 + L_w^6 p^6$

In the extreme case only the last term will be dominant, therefore the density of the final state can be written as:

$$\frac{dN}{dW} = \frac{4\pi p^2 dp V}{(2\pi\hbar)^3 L_w^6 p^6} \qquad \ldots (16)$$

This will give an additional term to the decay time given by equation (14), which is:

$$d\Gamma' = \frac{8g^2 |M|^2 p^7}{c^3 \hbar^7 m^2} \frac{1}{L_w^6 p^6} \qquad \ldots (17)$$

Equations (16) and (17) imply that the decay rate and consequently the decay time of the heavier states (particles) will have a very different energy (momentum) dependence as compared to the usual dependence given by equation (14).

Equation (14) would imply a $m^5$ dependence $\left(\sim p^7/m^2\right)$, whereas equations (16) and (17) imply a dependence $\left(\sim p/m^2\right)$, so in the case $E \gg \hbar c/L_w$, decay time goes as $m^{-1}$. In the region where $E$ is comparable to $\hbar c/L_w$, there would be a complicated dependence on $p$, as given by the expansion of equation (15) in powers of $L_w^2 p^2$.

For instance if two states differ in rest energy by a factor of 2, the decay rates would scale very differently in the situations described by equation (14) and equations (15) to (17). This would be a signature of the extended uncertainty principle at work.
The upper limit on the number density as:

$$n \leq \left(\frac{E}{\hbar c}\right)^3 \left(1 + 3L_w^2 p^2\right)^{-1} \quad \ldots (18)$$

And the constraint on the flux is given by:

$$f \leq \frac{E^4}{4c^2\hbar^3}\left(1 - \frac{3L_w^2}{\hbar^2}\left(\frac{E}{c}\right)^2\right) \quad \ldots (19)$$

Again equation (19) implies that the corrections to the flux, due to EUP, could vary by an order of magnitude. Also see reference [8].

**Reference:**

1. M Riordani, Physics World, p.16, Oct. 2008
2. N Arkani Ahmed et al, Phys. Lett. B, 429, 263, 1998; C Sivaram, Current Science, 79, 413, 2000
3. S B Giddings, Gen. Rel. Grav., 34, 1775, 2002; (For earlier discussion in another context see C Sivaram, Phys. Reports, 51, 111, 1979)
4. C Sivaram, Int. J. Theor. Phys., 26, 1127, 1987
5. See for example C Sivaram, Gen. Rel. Grav., 33, 1, 2001
6. L Xiang and Y Genshen, Mod. Phys. Lett., 20, 1823, 2005; (For earlier discussion see C Sivaram, Astrophys. Spc. Sci., 167, 335, 1990)
7. D H Perkins, Introduction to High Energy Physics, O.U.P, 1993
8. C Sivaram, Kenath Arun, C A Samartha, Mod. Phys. Lett A, 23, 1470, 2008